\documentclass[11pt]{article}
\usepackage{fullpage}
\usepackage[numbers]{natbib}
\usepackage{float}
\usepackage{amsmath}
\usepackage{url}

\usepackage{subcaption}
\usepackage[capitalize]{cleveref}

\usepackage{xspace}
\usepackage{tcolorbox}
\usepackage{tikz}
\usepackage{adjustbox}
\usepackage{array}
\usepackage{diagbox}

\newcolumntype{R}[2]{%
    >{\adjustbox{angle=#1,valign=B,raise=-5ex,lap=\width-(#2)}\bgroup}%
    l%
    <{\egroup}%
}
\newcommand*\rot{\multicolumn{1}{R{60}{1em}}}

\usepackage[normalem]{ulem}

\usepackage{enumitem}

\providecommand{\Kth}[1]{\ensuremath{{#1}^{\rm th}}}

\newcommand{\GraphCreation}{\textsc{Graph Creation}\xspace}
\newcommand{\IterativeAlgorithm}{\textsc{Iterative Algorithm}\xspace}
\newcommand{\Finalization}{\textsc{Finalization}\xspace}
\newcommand{\SelectPath}{\textsc{Select Path}\xspace}
\newcommand{\ChooseFlow}{\textsc{Choose Flow Amount}\xspace}
\newcommand{\UpdateResidualGraph}{\textsc{Update Residual Graph}\xspace}

\AtBeginDocument{%
  }

\begin{document}

\title{iFlow: An Interactive Max-Flow/Min-Cut Algorithms Visualizer}

\author{{\rm Muyang Ye, Tianrui Xia, Tianxin Zu, Qian Wang, and David Kempe}\\
\\ University of Southern California
}
\date{}
\maketitle

\begin{abstract}
  The Max-Flow/Min-Cut problem is a fundamental tool in graph theory, with applications in many domains, including data mining, image segmentation, transportation planning, and many types of assignment problems, in addition to being an essential building block for many other algorithms. 
  The Ford-Fulkerson Algorithm for Max-Flow/Min-Cut and its variants are therefore commonly taught in undergraduate and beginning graduate algorithms classes. However, these algorithms --- and in particular the so-called \emph{residual graphs} they utilize --- often pose significant challenges for students.

  To help students achieve a deeper understanding, we developed iFlow, an interactive visualization tool for the Ford-Fulkerson Algorithm and its variants. 
  iFlow lets users design or import flow networks, and execute the algorithm by hand. In particular, the user can select an augmentation path and amount, and then update the residual graph. The user is given detailed feedback on mistakes, and can also have iFlow auto-complete each step, to use it as a demonstration tool while still in the initial learning stages. iFlow has been made publicly available and open-sourced.
  
   We deployed iFlow in an undergraduate algorithms class, and collected students' self-reported learning benefits via an optional survey. All respondents considered the tool at least somewhat useful and engaging, with most rating it either as useful/engaging or very useful/engaging. Students also generally reported a significant increase in understanding of the algorithm.
\end{abstract}

\section{Introduction}
\label{sec:intro}

Maximum Flow and Minimum Cut are two central algorithmic problems in computer science.
For both problems, the input is a graph with \emph{capacities} on the edges as well as designated start and end nodes (called \emph{source} and \emph{sink}). 
The Maximum Flow problem asks how much flow (e.g., water or car traffic) can be routed from the source to the sink at steady state using the entire network; the Minimum Cut problem asks for the tightest ``bottleneck'' constricting flow from the source to the sink.
(Precise definitions are given in \cref{sec:max-flow-min-cut}.)

In addition to applications directly following the definition --- such as analyzing the capacity of a traffic network or egress flow through a building and their bottlenecks --- both problems have numerous less obvious applications \citep{ahuja:magnanti:orlin}. 
The maximum flow problem can be used to solve Maximum Bipartite Matching as well as many of its variants and generalizations; these problems occur frequently for many types of assignments of one type of items to another type when each individual item can only be assigned to a limited number of the other type (e.g., workers to jobs, students to classes, computations to machines, etc.).
In addition to being helpful for detecting bottlenecks in networks, Minimum Cut problems frequently arise when objects must be partitioned into two disjoint sets, subject to individual rewards/penalties for membership in one set vs.~the other and penalties for separating adjacent objects. Classic examples include image segmentation into two sets of pixels or partitions of projects into those that will be taken on and those that will not, subject to prerequisite constraints.

Both problems are solved by the Ford-Fulkerson Algorithm \citep{ford:fulkerson}.
Due to the problems' importance and the mathematical elegance of the algorithm and its analysis, Maximum Flow and Minimum Cut are frequently taught in undergraduate and beginning graduate algorithms classes in universities.
Unfortunately, they are often hard for students to grasp; in particular, the so-called \emph{residual graphs} and \emph{backwards edges} often cause difficulties for students. 

In general algorithms instruction, in order to help students with different learning styles cope with such difficulties \citep{yadav2016},
an important teaching approach is to use visual tools, since they often engage students \citep{stasko1993}. However, visual tools by themselves are usually not enough to lead to significant learning gains \citep{stasko1993,hundhausen2002,hundhausen2002meta}; 
based on an analysis of successful visual learning systems, \citet{jaime2009} hypothesize that in addition to visualization, successful systems also provide textual explanations, feedback on the students' actions, and advanced features \citep{jaime2009}.
Several other studies have drawn similar conclusions: to improve students' comprehension and retention of complex algorithms, several features are crucial. These include user control over visualization pace, focused logical steps, and the incorporation of interactive elements that actively engage learners \citep{saraiya2004}, as well as narrative content and textual explanations, integrating explanations with the steps of the visualization, feedback on student action, context-sensitive help, structural views of algorithms, and data input facilities \citep{8560314}.

Based on these insights into successful visual algorithms learning tools, we have developed iFlow, described in detail in \cref{sec:visualization}, publicly available at the iFlow website \citep{muyang:website}, and open-sourced on Github \citep{muyang:repository}. 
In the categorization of \citet{naps2002},
iFlow is a \emph{constructing} engagement level interactive visualization tool for Max-Flow and Min-Cut algorithms. It allows students to construct their own visualizations of the Ford-Fulkerson algorithm, thus making it \emph{interactive}. A very recent example of such interactive visualization is GILP \citep{robbins2023}, which visualizes the Simplex algorithm. iFlow enables students to create and import/export custom flow networks, choose augmenting paths, apply specific flow amounts, update residual graphs, and find minimum cuts themselves. Every step is accompanied by context-sensitive help and narrative instructions, detailed feedback if students make mistakes, and automatic completion (multi-ended for selecting augmenting paths).
These features serve as the ``guide'' for the student, making iFlow embody many features suggested by \citet{jaime2009, saraiya2004, 8560314} to be essential for successful algorithms visualization tools.
iFlow is suitable both as a demonstration tool for an instructor in a class setting and as a tool for students to practice independently.

We report (in \cref{sec:evaluation}) on experience with deploying iFlow in an undergraduate algorithms class at a major research university in the United States. In this class, iFlow was used both for classroom demonstration, and more importantly, as a tool for students to practice independently on one of the homework assignments. As part of this assignment, students were asked to optionally fill out a survey about their experience. The student feedback was overall positive, providing evidence both for the usefulness of iFlow and algorithm visualization more generally.

\subsection{Additional Related Work}
iFlow should be considered in the context of the development and depolyment of visual learning tools for algorithms instruction; this connection and motivation has been discussed in depth above.
Prior to our work, multiple Max-Flow visualization tools were available online; see \citet{heller2017, haucke2024, fischer2015, isabek2016} for representative examples.%
\footnote{In particular, the (open-sourced) tool of \citet{isabek2016} served as a point of departure for the development of the visualization component of iFlow.}
These tools allow users to run the Ford-Fulkerson algorithm and its variants on custom inputs. 
The main difference between these tools and iFlow is the lack of interactivity: while they \emph{show} the execution of the algorithm, they do not let users attempt to execute the steps by hand. 
For instance, even though users can step into the next line of illustrated pseudocode, see the updated flow network after executing that line, pause, and step back in \citet{haucke2024} and \citet{fischer2015}, users are not able to for example select an arbitrary augmenting path themselves; thus, it would be hard for users to learn the difference between various strategies to select augmenting paths by the visualizations alone. 
Because users have no autonomy over the execution of the algorithm, in particular, they can never be given feedback on their actions. In addition, existing tools typically lack narrative instructions on each step and do not visualize minimum cuts.
\section{Max-Flow/Min-Cut: Definitions and Algorithms} 
\label{sec:max-flow-min-cut}

\subsection{Problem Definition}
The maximum $s$-$t$ flow problem is a fundamental algorithm problem \citep{ahuja:magnanti:orlin,kleinberg:tardos:algorithm-design}, defined as follows. 
The input is a graph $G=(V,E)$ in which each edge $e$ has a \emph{capacity} $c_e \geq 0$, together with a designated \emph{source node} $s \in V$ and \emph{sink node} $t \in V$. 
A \emph{flow} $f$ is an assignment of values $f_e$ to all edges $e$, satisfying non-negativity ($f_e \geq 0$ for all $e$), capacity constraints ($f_e \leq c_e$ for all $e$), and conservation ($\sum_{e \text{ into } v} f_e = \sum_{e \text{ out of } v} f_e$ for all nodes $v \neq s, t$).
The \emph{value} of the flow $f$ is $\nu(f) = \sum_{e \text{ out of } s} f_e$; it is easy to show that this equals $\sum_{e \text{ into } t} f_e$.
A \emph{maximum $s$-$t$ flow} is a flow maximizing $\nu(f)$.

An \emph{$s$-$t$ cut} $(S, \overline{S})$ is a partition of the vertices $V$ into sets $S, \overline{S}$ such that $s \in S$ and $t \in \overline{S}$.
The \emph{capacity} of the cut $(S, \overline{S})$ is the total capacity of all edges $e=(u,v) \in E$ with $u \in S$ and $v \in \overline{S}$, i.e., $c(S, \overline{S}) = \sum_{e=(u,v) \in E, u \in S, v \in \overline{S}} c_e$.
The \emph{minimum $s$-$t$ cut problem} consists of finding a cut $(S,\overline{S})$ minimizing $c(S,\overline{S})$.
A useful fact about minimum $s$-$t$ cuts is that they are closed under intersections and unions; that is, if $(S_1, \overline{S_1})$ and $(S_2, \overline{S_2})$ are minimum $s$-$t$ cuts, then so are $(S_1 \cap S_2, \overline{S_1 \cap S_2})$ and $(S_1 \cup S_2, \overline{S_1 \cup S_2})$. In particular, this implies that the notion of the \emph{smallest} minimum $s$-$t$ cut is well-defined, i.e., there is a unique minimum $s$-$t$ cut minimizing $|S|$.

\subsection{Algorithms}
Due to the well-known duality between flows and cuts, once one has a maximum $s$-$t$ flow, one can find a minimum $s$-$t$ cut with a simple Breadth-First Search (BFS). 
Specifically, starting from the source $s$, and following all edges $e$ with $f_e < c_e$ (such edges are called \emph{non-saturated}), yields a set $S$ with $s \in S$ and $t \in \overline{S}$, and it is well-known that this set gives a minimum $s$-$t$ cut \citep{ahuja:magnanti:orlin,kleinberg:tardos:algorithm-design}.

The \emph{Ford-Fulkerson Algorithm} \citep{ford:fulkerson} is the most well-known method for computing the maximum $s$-$t$ flow (and with it a minimum $s$-$t$ cut). It is based on the central concept of a \emph{residual graph}. The residual graph $G_f$ for a given flow $f$ contains, for each original edge $e=(u,v)$, the edge $(u,v)$ with \emph{residual capacity} $c'_e = c_e - f_e$, and, importantly, the \emph{backwards edge} $(v,u)$ with residual capacity $c'_{(v,u)} = f_e$, capturing the fact that up to $f_e$ units of flow can be ``undone'' on $e=(u,v)$ by pretending to send flow from $v$ to $u$ instead.
In each iteration, the Ford-Fulkerson Algorithm finds a path $P$ from $s$ to $t$ in $G_f$ such that each edge $e \in P$ in the path has positive residual capacity $c'_e > 0$. Such a path is called an \emph{augmenting path} for the flow. The algorithm then adds as much flow as possible on this path; the maximum is the \emph{bottleneck capacity} $b = \min_{e \in P} c'_e$. For each edge $e \in P$, if $e$ is a regular edge (goes forward), then the flow is updated to $f'_e = f_e + b$. If $(v,u) \in P$ is a backwards edge, then instead $b$ is \emph{subtracted} from the flow on the edge $(u,v)$, i.e., $f'_{(u,v)} = f_{(u,v)} - b$. 
After the flow is updated, so is the residual graph.
This loop --- find an augmenting path, add as much flow as possible, update the residual graph --- is repeated until no more augmenting path $P$ from $s$ to $t$ exists, at which point the algorithm terminates, and a maximum $s$-$t$ flow has been found.

When all capacities are integer, the running time of the algorithm is upper-bounded by $O(|E| \cdot \nu(f^*))$, where $f^*$ is the flow at termination.
However, if the augmenting paths are chosen in no particular way, the running time could in fact be as large as this bound, which could possibly be very large (if the network has large capacities)
Two variants of the algorithm achieve better running time for such inputs.

The \emph{Edmonds-Karp Algorithm} \citep{edmonds:karp} is a specific implementation of the Ford-Fulkerson Algorithm which always uses a \emph{shortest} (fewest edges) augmenting path (which can be found in time $O(|V|+|E|)$ using breadth-first search). This choice guarantees that the algorithm finishes in time $O(|V| \cdot |E|^2)$, regardless of how large the capacities are.

The \emph{Widest Path Heuristic} is another variant of the Ford-Fulkerson Algorithm. It chooses, in each iteration, an augmenting path $P$ with largest possible bottleneck capacity $b$.\footnote{Such a path can be found in polynomial time, e.g., with variants of Dijkstra's shortest-path algorithm, or binary search and breadth-first search.}
Using the Widest Path Heuristic results in running time $O(|E|^2 \cdot \log (|E|) \cdot \log \nu(f^*))$, exponentially improving the dependency on the worrisome quantity $\nu(f^*)$ (which could be very large), as also shown by \citet{edmonds:karp}.

\section{The Interactive Visualization}
\label{sec:visualization}

iFlow is structured into three distinct stages: 
(1) \GraphCreation, (2) \IterativeAlgorithm, 
and (3) \Finalization. Within the \IterativeAlgorithm stage, there are three phases: (2.1) \SelectPath, (2.2) \ChooseFlow, and (2.3) \UpdateResidualGraph. In the different stages, iFlow has the same layout, as shown in \cref{plot:figure2} using \SelectPath  as an example, but different functionalities (i.e., different buttons and allowed operations on the graph) in every stage. There are three areas in \cref{plot:figure2} horizontally. The leftmost area displays the instructions for the current stage. The middle area contains the (residual) graph itself, and users are able to interact with the graph. 
Even though the residual graph by itself is enough for performing an iteration of the Ford-Fulkerson Algorithm, we have provided the option to show or hide the original capacities and current flow (``applied flow/original capacity''), as shown by the blue edges in \cref{plot:figure2}, to help students better understand the relationship between the flow network and its residual graph. The rightmost area records the history of augmenting paths chosen and the flow amount added on those paths so that students always have a record of how the current flow was obtained. The rightmost area also contains three buttons. After a maximum flow is found, students can utilize those buttons to confirm the final flow amount, verify a proposed minimum cut, or let the tool find a minimum cut.

iFlow starts in the \GraphCreation stage. Once the user has confirmed the graph, source, and sink, they start the interactive execution of the algorithm, which lasts until the user determines that there is no more augmenting path. The following sections delve into the detailed functionalities of each stage.

\subsection{Stage 1: \GraphCreation}
\begin{figure}[htb]
  \centering
  \includegraphics[width=\columnwidth]{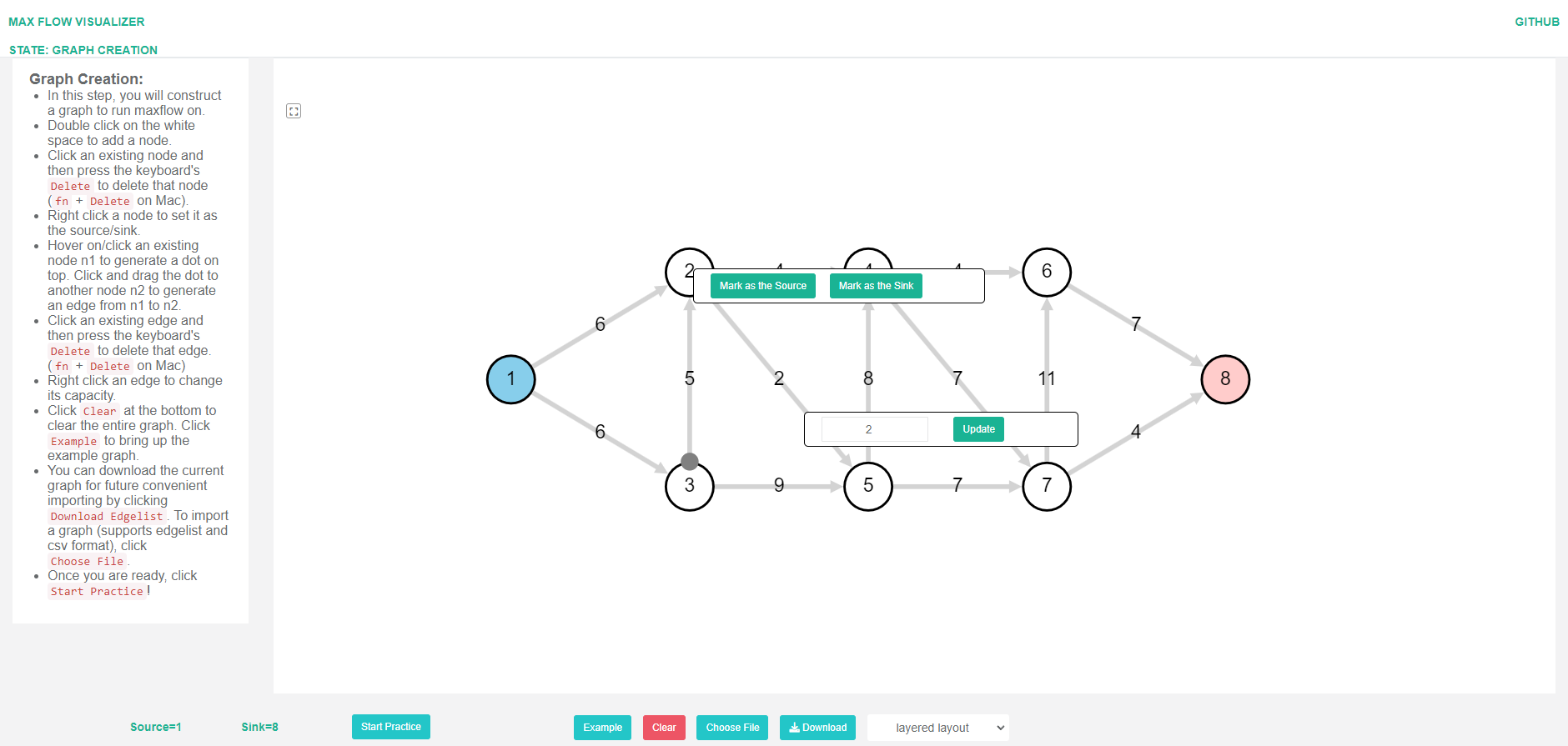}
  \caption{\GraphCreation Stage}
  \label{plot:figure1}
\end{figure}

In this stage, as illustrated in \cref{plot:figure1}, the user constructs a flow network. Specifically, they can add/delete nodes/edges, mark a node as the source or sink, modify edges' capacities, change the layout of the graph, and import/export the flow network in edgelist format.

\subsection{Stage 2: \IterativeAlgorithm}
In the second stage, the student executes the Ford-Fulkerson Algorithm iteratively by hand. Following the outline provided in \cref{sec:max-flow-min-cut}, in each iteration, the student first selects a path on which to add flow, then chooses the amount of flow to add, and finally updates the residual capacities of the edges.

\subsubsection{Phase 2.1: \SelectPath }
\begin{figure}[htb]
  \centering
  \includegraphics[width=\columnwidth]{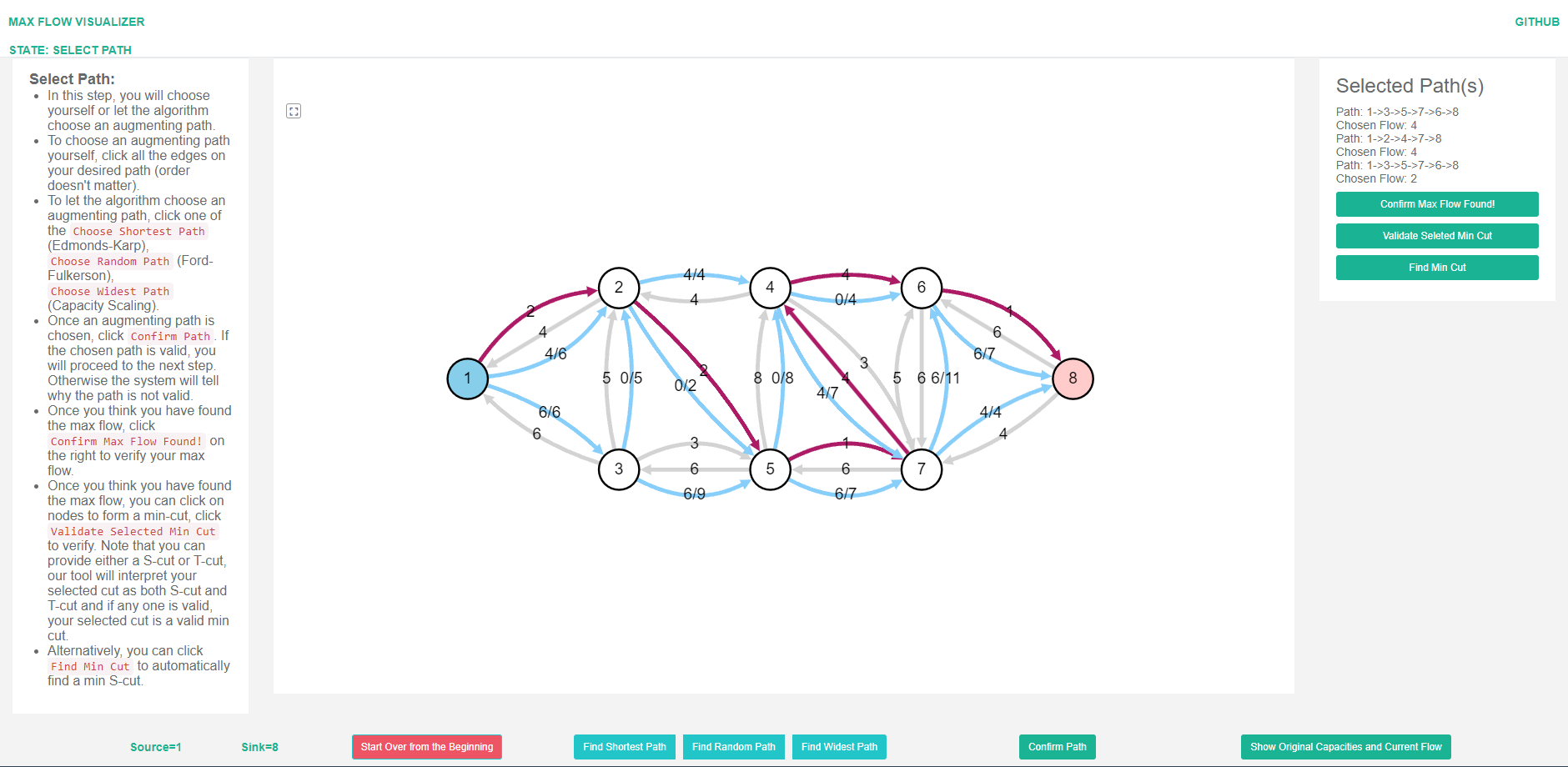}
  \caption{\SelectPath Stage}
  \label{plot:figure2}
\end{figure}

First, students select an augmenting path by clicking edges; selected edges are highlighted. \cref{plot:figure2} shows an example of a selected path containing a backward edge.

Students then are able to enter the next phase, once iFlow validates that the selected path has valid topology: starting from the source and ending at the sink with no redundant edges. 
Alternatively, the user may let iFlow automatically find a random\footnote{More precisely, a depth-first search starting from the source such that each outgoing edge of the current node that leads to an unvisited node is equally likely to be chosen.} augmenting path, a shortest augmenting path, or a widest augmenting path; the latter two correspond to the Edmonds-Karp and Widest Path versions. 
This feature serves three purposes. 
First, it allows iFlow to be used as a non-interactive demonstration tool, e.g., in a classroom setting. 
Second, in case a student does not understand the concept of augmenting paths yet, it provides an example. 
Finally, after the student is proficient in finding augmenting paths, it accelerates the process so that students can focus on practicing other topics such as finding the residual graph or a minimum cut.

\subsubsection{Phase 2.2: \ChooseFlow}
\begin{figure}[htb]
  \centering
  \includegraphics[width=\columnwidth]{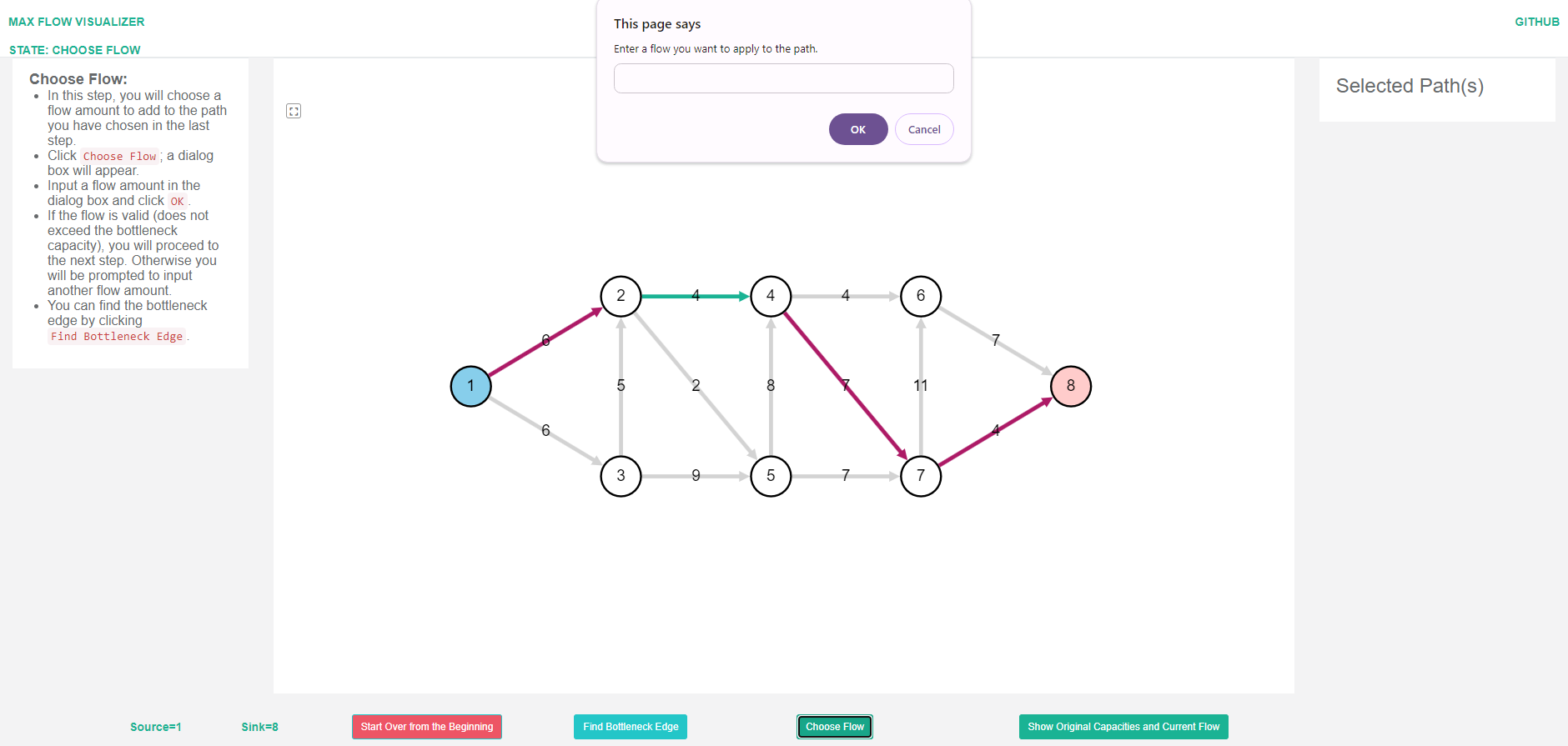}
  \caption{\ChooseFlow Stage}
  \label{plot:figure3}
\end{figure}

\cref{plot:figure3} demonstrates the \ChooseFlow phase. The purple-colored edges 
show the augmenting path chosen in the previous phase. Students can decide on a flow amount to be applied on this path. 
The bottleneck capacity (denoted by $b$ in \cref{sec:max-flow-min-cut}) is a particularly natural choice; therefore, iFlow provides a button to highlight the bottleneck edge along the chosen path. Students can apply less flow if they want to experiment. If the entered flow amount is greater than the bottleneck residual capacity on this path, an error message is shown.

\subsubsection{Phase 2.3: \UpdateResidualGraph}
\begin{figure}[htb]
  \centering
  \includegraphics[width=\columnwidth]{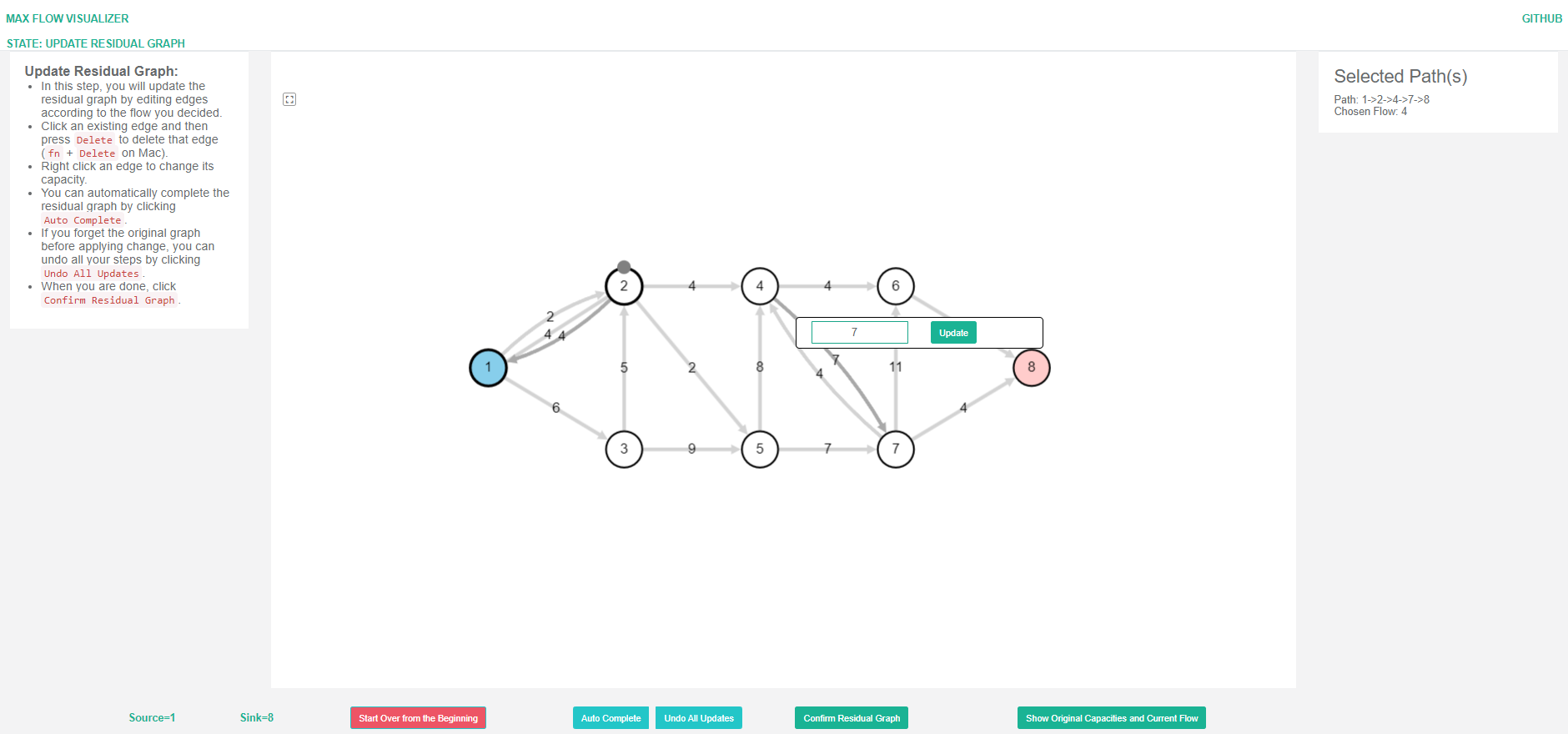}
  \caption{\UpdateResidualGraph Stage}
  \label{plot:figure4}
\end{figure}

In the \UpdateResidualGraph phase, students construct the updated residual graph after selecting an augmenting path and flow amount; this step is often one of the most difficult concepts for students to grasp. 
In \cref{plot:figure4}, a user is attempting to add a backward edge and modify the capacity of an edge along the previously chosen augmenting path. Like in all other phases, there is a convenient button \textit{Auto Complete} to automatically construct the residual graph. Similar to the rationale behind automatically finding an augmenting path, this feature serves multiple purposes: (1) it exemplifies the concept of residual capacities when students are not ready yet, and (2) it avoids tedious repetition when users are already very comfortable with updating residual capacities. In particular, it allows an instructor to use iFlow non-interactively in a classroom setting. If students choose to manually update the residual graph and make mistakes, iFlow will display a detailed error message (e.g., which edges have the wrong residual capacity). 

\subsection{Stage 3: \Finalization}
Finally, if at any point, the student believes that the flow is maximum, they can ask iFlow to confirm whether it is indeed. If there is still an augmenting path, the student is given an error message and instructed to continue finding augmenting paths. Otherwise, by clicking \textit{Confirm Max Flow Found!}, the student is prompted to enter the value of the flow; iFlow will check whether the flow is indeed a maximum flow for the given flow network. Asking students to specify the value tests whether they understand the concept of the value of a flow.

\begin{figure}[htb]
  \centering
  \includegraphics[width=\columnwidth]{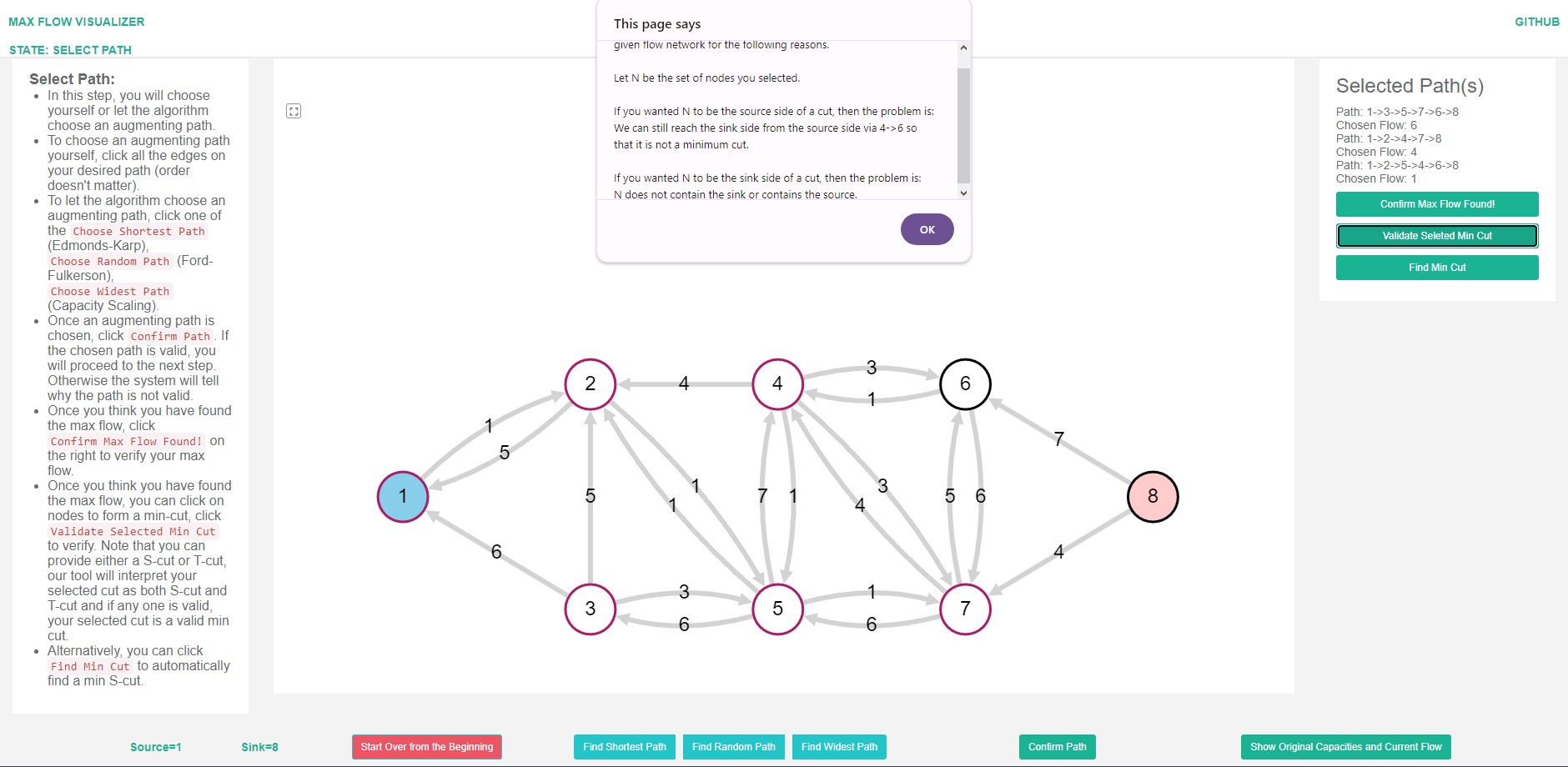}
  \caption{An Example Min-Cut Error Message} 
  \label{plot:figure5}
\end{figure}

In addition to confirming a maximum flow, students can click nodes to mark them as the nodes on one side of a minimum cut (highlighted in purple), and then click \textit{Validate Selected Min Cut} to verify the proposed cut. iFlow will interpret the selected nodes as the $s$-side of a cut or the $t$-side of a cut. If the selected nodes are the $s$-side or $t$-side of any minimum $s$-$t$ cut, students are congratulated for finding a minimum cut. Otherwise, they will be shown a detailed error message that points out why the selected nodes do not constitute either the $s$-side or $t$-side of a minimum $s$-$t$ cut; this is shown in \cref{plot:figure5}. 
Alternatively, students can simply click \textit{Find Min Cut} to obtain the $s$-$t$ cut $(S, \overline{S})$ minimizing $|S|$. 

\subsection{Implementation}
iFlow is a pure client side application since there is no storage needed except possibly for the graph in edgelist format; this small amount of data is much more naturally stored locally. Our iFlow implementation contains code in HTML, CSS, and JavaScript; all code has been open-sourced on Github \citep{muyang:repository} under the MIT license.
The Cytoscape framework is used for graph visualization and interaction. The graph data model and all algorithms such as finding augmenting paths and verification of a minimum cut are implemented with native JavaScript from scratch. 
The code is structured such that the graph data model/algorithms and the interactive visualization are decoupled, making it easily extensible.
On the front-end, we have employed Bootstrap for a grid-based horizontal page layout. We have also drawn heavily on the open-sourced non-interactive visualization MAX FLOW AND MIN CUT \citep{isabek2016} for the user interface. Specifically, we have kept the same vertical layout and retained most of Cytoscape's settings such as nodes' size from that repository.

In order to improve legibility, in particular for large graphs, we have implemented from scratch two types of graph layout: a Spring Model Layout and a Layered Layout. Both try to place nodes to avoid overlap and improve usability. The addition of future layout algorithms would be easy. Users can also manually adjust node positions if the default layout does not meet their needs. 
When a graph is uploaded from an edgelist file, if the file contains position information for the nodes, the nodes will be positioned according to this information.
\section{Evaluation}
\label{sec:evaluation}

To evaluate the impact of iFlow on students' (self-reported) understanding of Max-Flow/Min-Cut algorithms, and also simply to have students benefit from using the tool, we included a question requiring the use of iFlow in a graded homework assignment in an undergraduate algorithms class at the University of Southern California, a major research university in the United States.

\subsection{Setup}
\label{sec:evaluation-setup}
The class in which iFlow was used is taken mostly by undergraduate students in their second or third year of study. During the semester on which we report, 162 students were enrolled, of whom 124 submitted this particular homework.
The specific question asked students to use iFlow to simulate the Ford-Fulkerson Algorithm on a flow network derived from a Maximum Bipartite Matching instance, and specified a tie-breaking rule for path finding which would ensure that the algorithm would frequently use backwards edges and add/remove flow on the same edges.
Students were asked to optionally fill out a questionnaire about their experience with iFlow. This questionnaire was filled out by 21 students\footnote{31 students intially started the questionnaire, but 10 did not fill in any answers.}. The full text of both the homework question related to iFlow and the questionnaire is included in the appendix.

The questionnaire served two purposes: (1) to obtain feedback on bugs, unclear instructions, etc. Based on the responses, iFlow has already been improved to address all the issues that were raised. 
(2) To evaluate to what extent iFlow improved the students' (self-reported) understanding of the Ford-Fulkerson Algorithm. Here, we report on the outcomes for questions related to the learning experience.
In addition to one open-ended free-text format question, we designed four questions, evaluated on a 1--5 Likert scale\footnote{For RQ1, 1=not at all engaging, 2=a little engaging, 3=reasonably engaging, 4=engaging, 5=very engaging. For RQ2--4, 1=not at all useful, 2=a little useful, 3=reasonably useful, 4=useful, 5=very useful.}.
We specifically included a question about student engagement, because several studies have demonstrated that student engagement levels are a crucial indicator of the effectiveness of algorithm visualization tools in enhancing learning outcomes in Computer Science Education \citep{hundhausen2000, naps2002, kehoe2001, schweitzer2007}. 
The four questions are as follows:

\begin{description}
    \item[RQ1:] How engaging did the tool make learning/practicing the Max-Flow Min-Cut Algorithm?
    \item[RQ2:] How useful was the tool's visualization to help you understand the Max-Flow Min-Cut Algorithm?
    \item[RQ3:] How much did the tool's self-test feature help you understand the Max-Flow Min-Cut Algorithm?
    \item[RQ4:] How useful was the feedback when you made a mistake in your solution?
\end{description}

\subsection{Results}
\label{sec:evaluation-results}

\begin{figure}[H]
  \centering
  \includegraphics[width=\columnwidth]{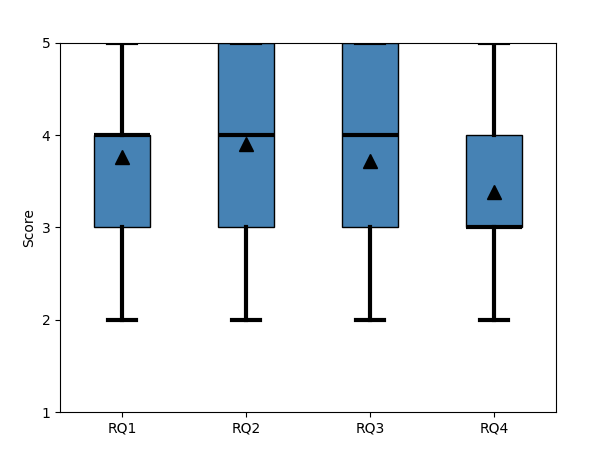}
  \caption{Aggregated opinion scores for four research questions. Thick line=median, triangle=mean, box lower end=\Kth{25} percentile, box upper end=\Kth{75} percentile, lines extending from box=range (max to min)}
  \label{plot:aggregate}
\end{figure}

\cref{plot:aggregate} shows the distribution of students' responses for the four main questions. The average opinion score is 3.76 for RQ1, 3.90 for RQ2, 3.71 for RQ3, and 3.38 for RQ4. For all research questions, no response rated iFlow most negative. For RQ2 and RQ3, over 25\% of students expressed that the ability to visualize the flow network and interactively self-test is very useful.

Beyond analyzing raw opinion scores, we are interested in whether iFlow is more effective for students with a specific level of prior understanding of the topic. For example, is iFlow helpful for students to understand the algorithm for the first time? Does it solidify partial understanding? Do advanced students also find iFlow useful?
To enable such a more fine-grained analysis, we included a question asking students to rate how well they understood the Ford-Fulkerson Algorithm before using iFlow on a 5-point Likert scale\footnote{1=not at all; 2=a little bit; 3=reasonably well; 4=fairly well; 5=very well.}; we then calculated the number of responses to RQ1--4 conditioned on the answers to this question, as shown in \cref{tab:engaging,tab:visualization,tab:self-test,tab:feedback}.

\begin{table}[H]
\centering
\caption{Number of RQ1 (“How engaging did the tool make learning/practicing the Max-Flow Min-Cut Algorithm?”) responses given self-reported prior understanding of Max-Flow and Min-Cut algorithms \label{tab:engaging}}
\resizebox{0.7\textwidth}{!}{%
\begin{tabular}{ | c | c | c | c | c | c |} 
     \multicolumn{1}{c}{\diagbox[height=15ex]{Prior\\Understanding}{How engaging\\was iFlow?}}
     & \rot{not at all engaging} 
     & \rot{a little engaging} 
     & \rot{reasonably engaging} 
     & \rot{engaging} 
     & \rot{very engaging} 
     \\  \hline
    not at all & 0 & 0 & 0 & 2 & 1 \\ \hline
    a little bit & 0 & 0 & 2 & 3 & 2 \\ \hline
    reasonably well & 0 & 1 & 2 & 2 & 1 \\ \hline
    fairly well & 0 & 0 & 3 & 0 & 0 \\ \hline
    very well & 0 & 0 & 0 & 1 & 1 \\ \hline
\end{tabular}
}
\end{table}

\begin{table}[H]
\centering
\caption{Number of RQ2 (“How useful was the tool’s visualization to help you understand the Max-Flow Min-Cut Algorithm?”) responses given self-reported prior understanding of Max-Flow and Min-Cut algorithms \label{tab:visualization}}
\resizebox{0.7\textwidth}{!}{%
\begin{tabular}{ | c | c | c | c | c | c |} 
     \multicolumn{1}{c}{\diagbox[height=15ex]{Prior\\Understanding}{Visualization\\Usefulness}}
     & \rot{not at all useful} 
     & \rot{a little useful} 
     & \rot{reasonably useful} 
     & \rot{useful} 
     & \rot{very useful} 
     \\ \hline
    not at all & 0 & 1 & 1 & 1 & 0 \\ \hline
    a little bit & 0 & 0 & 2 & 2 & 3 \\ \hline
    reasonably well & 0 & 1 & 1 & 0 & 4 \\ \hline
    fairly well & 0 & 0 & 1 & 2 & 0 \\ \hline
    very well & 0 & 0 & 1 & 0 & 1 \\ \hline
\end{tabular}
}
\end{table}

\begin{table}[H]
\centering
\caption{Number of RQ3 (“How much did the tool’s self-test feature help you understand the Max-Flow Min-Cut Algorithm?”) responses given self-reported prior understanding of Max-Flow and Min-Cut algorithms \label{tab:self-test}
}
\resizebox{0.7\textwidth}{!}{%
\begin{tabular}{ | c | c | c | c | c | c |} 
     \multicolumn{1}{c}{\diagbox[height=15ex]{Prior\\Understanding}{Self-Test\\Usefulness}}
     & \rot{not at all useful} 
     & \rot{a little useful} 
     & \rot{reasonably useful} 
     & \rot{useful} 
     & \rot{very useful} 
     \\ \hline
    not at all & 0 & 1 & 1 & 0 & 1 \\ \hline
    a little bit & 0 & 0 & 2 & 3 & 2 \\ \hline
    reasonably well & 0 & 1 & 2 & 1 & 2 \\ \hline
    fairly well & 0 & 0 & 1 & 2 & 0 \\ \hline
    very well & 0 & 1 & 0 & 0 & 1 \\ \hline
\end{tabular}
}
\end{table}

\begin{table}[H]
\centering
\caption{Number of RQ4 (“How useful was the feedback when you made a mistake in your solution?”) responses given self-reported prior understanding of Max-Flow and Min-Cut algorithms \label{tab:feedback}
}
\resizebox{0.7\textwidth}{!}{%
\begin{tabular}{ | c | c | c | c | c | c |} 
     \multicolumn{1}{c}{\diagbox[height=15ex]{Prior\\Understanding}{Feedback\\Usefulness}}
     & \rot{not at all useful} 
     & \rot{a little useful}  
     & \rot{reasonably useful} 
     & \rot{useful} 
     & \rot{very useful} \\ \hline
    not at all & 0 & 1 & 1 & 0 & 1 \\ \hline
    a little bit & 0 & 0 & 2 & 3 & 2 \\ \hline
    reasonably well & 0 & 2 & 3 & 1 & 0 \\ \hline
    fairly well & 0 & 1 & 0 & 2 & 0 \\ \hline
    very well & 0 & 0 & 2 & 0 & 0 \\ \hline
\end{tabular}
}
\end{table}

For all questions, there is no clear correlation between responses and self-rated previous understanding ($r=0.074$ for RQ1, $r=0.138$ for RQ2, $r=-0.028$ for RQ3, $r=-0.247$ for RQ4). However, we have observed two minor results. First, students who answered that they did not know max-flow at all found iFlow to be either engaging or very engaging. Second, as students' self-rated prior understanding increases, the maximum level of perceived usefulness of mistake feedback decreases. One conjecture is that more advanced students simply did not make many mistakes, so they got little to no feedback. 
Again, due to low absolute $r$-values, it is possible that these two observations are entirely due to the relatively small sample size, although they appear intuitively plausible.

The last question of our questionnaire let students share any feedback/comments/suggestions. This question received only six responses, which are included in their entirety in the appendix. Four of these responses were concerned with bugs or feature requests. The two responses ``evaluating'' iFlow expressed that iFlow was very useful. In particular, one stated: ``[...] I tried to spend the day reading and watching tutorials but non of them make me understand the concept well. But, after I played around this tool I instantly get how everything works, to a level impossible by reading alone. [...]'' [sic]
The main complaints were about confusing buttons and that the mistake feedback could be more specific. For the former, there were one or two buttons that broke or had insufficient instructions for different operating systems initially; we have addressed all of these concerns since the feedback was collected.
\section{Conclusions and Future Work}

We presented iFlow, an interactive tool for visualizing Ford-Fulkerson-type Max-Flow/Min-Cut algorithms. Students can practice finding augmenting paths, applying flows, constructing residual graphs, and finding minimum cuts on their own in distinct stages. Our survey shows that iFlow makes learning Max-Flow/Min-Cut algorithms engaging, and students believed that its features of visualization, self-test, and mistake feedback are generally useful. 

There are several directions for future work. Naturally, one could implement other Max-Flow/Min-Cut algorithms, such as Dinic's Blocking Flows Algorithm or Preflow-Push based algorithms. We consider these algorithms more advanced, and --- different from the Ford-Fulkerson Algorithm --- they are rarely taught in introductory undergraduate classes. Thus, the course audience when these algorithms are taught might be in less need of a visualization tool.

Another natural next step would be a more large-scale evaluation of iFlow. This would include larger classes (to obtain a higher number of responses, and thus more statistical significance), as well as different schools and class levels (undergraduate and masters students).
As with most educational innovations, a ``holy grail'' would be a proper randomized A/B test to evaluate students' actual learning gains, rather than merely self-reported benefits. As with almost all settings, such an A/B test would be met with several challenges: (1) the ethical challenge of depriving a large student population of a possibly useful learning tool which their peers may have access to, and (2) the typical ``leakage'' problem under which students in the control group obtain access to the tool through friends in the treatment group.
\section{Acknowledgments}
\label{sec:acknowledgments}
  We thank Gisele Ragusa for useful discussions and advice, and Matthew Ferland for userful pointers to the literature. We also thank Isabek Tashiev for open-sourcing the code for MAX FLOW AND MIN CUT. Finally, we thank students for providing valuable feedback on iFlow.

\clearpage 

\bibliographystyle{ACM-Reference-Format}
\bibliography{references}

\newpage

\appendix

\section{The Homework Problem Utilizing iFlow}
\label{sec:appendix-homework}

We give the precise text of the homework question the students were asked to tackle using iFlow.

\begin{tcolorbox}[height=0.85\textheight]

\vspace{1ex}

    \begin{center}
\begin{tikzpicture}
  [auto,
  defaultnode/.style={circle,draw=black}]

  \pgfsetxvec{\pgfpoint{0.8cm}{0cm}}
  \pgfsetyvec{\pgfpoint{0cm}{0.8cm}}

  \node[defaultnode] (a) at (0,6) {A};
  \node[defaultnode] (b) at (0.2,4) {B};
  \node[defaultnode] (c) at (0.4,2) {C};
  \node[defaultnode] (d) at (0.6,0) {D};
  \node[defaultnode] (e) at (5,6) {E};
  \node[defaultnode] (f) at (4.8,4) {F};
  \node[defaultnode] (g) at (4.6,2) {G};
  \node[defaultnode] (h) at (4.4,0) {H};

  \draw [line width = 0.5pt,-] (a) to (e);
  \draw [line width = 0.5pt,-] (a) to (f);
  \draw [line width = 0.5pt,-] (a) to (g);
  \draw [line width = 0.5pt,-] (a) to (h);
  \draw [line width = 0.5pt,-] (b) to (e);
  \draw [line width = 0.5pt,-] (b) to (f);
  \draw [line width = 0.5pt,-] (b) to (g);
  \draw [line width = 0.5pt,-] (c) to (e);
  \draw [line width = 0.5pt,-] (c) to (f);
  \draw [line width = 0.5pt,-] (d) to (e);

\draw (2.5, -1) node {\small Figure 1: An example input graph for bipartite matching};


\end{tikzpicture}
\end{center}

\small 
Consider the example bipartite graph shown in Figure 1.
Suppose you want to find a maximum matching in this graph. You can probably find one very quickly by hand, and will notice that it is a perfect matching.
Here, you are supposed to practice your understanding of the Ford-Fulkerson Max-Flow Algorithm by applying the reduction from class (from \textsc{Bipartite Matching} to \textsc{Maximum Flow}), then running the Ford-Fulkerson Algorithm by hand, with a specific tie breaking rule.
You are supposed to do the latter with a tool that a group of students
wrote to help you understand the Ford-Fulkerson Algorithm better.
The tool is at 
\url{https://maxflow-visualization.github.io/iFlow/}.

\begin{enumerate}
    \item {[2 points]} Using a screenshot from the tool, give the $s$-$t$ flow network that comes out of the reduction.
    \item {[6 points]} Run the Ford-Fulkerson Algorithm in self testing mode, where you find the paths and update the residual capacities yourself.
    Specifically, since in general, there may be many augmenting paths, you are supposed to use the following tie breaking rule for picking an augmenting path. Among all augmenting paths, pick the one whose first node is earliest in the alphabet. If there are multiple paths, pick the one whose second node is earliest in the alphabet. If there are multiple such paths, pick the one whose third node is earliest in the alphabet. And so on. (This would be called the ``lexicographically smallest'' path.)

    After each iteration, the flow you have can be interpreted as a matching (not necessarily a maximum one until the end).
    Describe what the matching is after each iteration. Ideally, this would be a figure, drawn by hand or screen-shotted. But if that's too much work, you can list all the edges in the matching.
    Describe in words what the algorithm is doing in terms of matching assignments on this input.

\item {[0 points]} Please share your feedback on the experience with the visualization tool at 
\textbf{[URL redacted]}.
\end{enumerate}
\end{tcolorbox}
\section{The Full Questionnaire}
\label{sec:appendix-questionnaire}

Here, we give the full questionnaire described in \cref{sec:evaluation-setup}, omitting the consent form and the eligibility question.

\begin{enumerate}[align=left]
    \item [\textbf{True/False:}] Did you encounter any bugs or highly unexpected behaviors in using the Max-Flow Visualization tool?
    \item [\textbf{(Optional) Free Response:}] If yes, please describe the bugs or unexpected behaviors.
    \item [\textbf{True/False:}] Did you encounter any difficulties other than bugs using the tool (e.g., unclear instructions, unintuitive interface, etc.)?
    \item [\textbf{(Optional) Free Response:}] If yes, what were the difficulties you encountered?
    \item [\textbf{Likert Scale:}] Did you find the instructions accurate and helpful?
    \item [\textbf{Likert Scale:}] How engaging did the tool make learning/practicing the Max-Flow Min-Cut Algorithm?
    \item [\textbf{Likert Scale:}] How well did you feel you understood the Max-Flow Min-Cut Algorithm before using the tool?
    \item [\textbf{Likert Scale:}] How useful was the tool's visualization to help you understand the Max-Flow Min-Cut Algorithm?
    \item [\textbf{Likert Scale:}] How much did the tool's self-test feature help you understand the Max-Flow Min-Cut Algorithm?
    \item [\textbf{Likert Scale:}] How useful was the feedback when you made a mistake in your solution?
    \item [\textbf{(Optional) Free Response:}] Any additional feedback, comments, or suggestions? In particular, if your experience on any of the earlier points was negative, please share details.
\end{enumerate}
\section{All Answers for the Last Question in the Questionnaire}
\label{sec:appendix-all-answers}

Here, we give all responses to the last question in \cref{sec:appendix-questionnaire}, which is also referred to in \cref{sec:evaluation-results}.

\begin{itemize}[leftmargin=*]
    \item ``I think it would be good to make it easier to delete edges''
    \item ``I feel like some functionality that is assigned to key presses such as deleting a node should be given interface buttons or selections. I had to read the instructions to understand how to delete a node when everything else was intuitive. It would also be cool if the tool displayed how much flow you have currently going from the source to the sink as you build your solution.''
    \item ``This is an extremely helpful visualization tool. It is far better than 5 hours of reading. I tried to spend the day reading and watching tutorials but non of them make me understand the concept well. But, after I played around this tool I instantly get how everything works, to a level impossible by reading alone. I know some students can get the concept just by reading it but unfortunately I am not one of those students. I learn the best by hands-on experience. I've noticed this difference in learning styles since high school. Thank you for developing such an amazing tool!''
    \item ``Overall helpful in many ways, other than being a little confusing!''
    \item ``I think the feedback could be more useful if it were more specific.''
    \item ``I think the button that automatically aligns the nodes doesn't make it clear that its purpose is to automatically align the nodes. It looks more like it's supposed to enlarge the screen''
\end{itemize}

\end{document}